\documentclass[11pt]{article}

\def\be{\begin{equation}}
\def\ee{\end{equation}}
\def\bea{\begin{eqnarray}}
\def\eea{\end{eqnarray}}
\newcommand{\nn}{\nonumber}
\newcommand{\de}{\partial}
\def \Mp {M_{\rm P}}
\def \rarr {\rightarrow}

\def \l {\lambda}
\def \L {\Lambda}

\def \b {\beta}
\def \a {\alpha}

\def \g {\gamma}

\def \d {\delta}

\def \psih {\hat{\psi}}

\begin{document}

\begin{titlepage}

\begin{flushright}
BA-TH/05-529\\
hep-th/0511147
\end{flushright}

\begin{center}

\Huge {Relic gravitons on Kasner branes}

\vspace{1cm}

\large{G. De Risi$^{1,2}$}

\normalsize \vspace{.2in}

{\sl $^1$Dipartimento di Fisica , Universit\`a di Bari, \\
Via G. Amendola 173, 70126 Bari, Italy}

\vspace{.2in}

{\sl $^2$Istituto Nazionale di Fisica Nucleare, Sezione di Bari\\
Via G. Amendola 173, 70126 Bari, Italy}

\vspace*{1.5cm}

\begin{abstract}
We present a simple model in which a brane with anisotropic metric is embedded in an AdS bulk.
We discuss the localization of the massless mode and the amplification of both the massless and the
massive modes on the branes, paying particular attention to the normalization of the perturbed action
and to the evaluation of the effective coupling constant that controls the amplitude of the spectrum.
In the model under investigation there is no mass gap between massless and massive modes,
and the massive modes can be amplified, with  mass-dependent amplitudes.

\end{abstract}
\end{center}

\end{titlepage}
\newpage

The possibility that our universe has more than 4 dimensions, proposed a long time ago, is now
widely investigated, since string theory needs 10 (or 11) dimensions to be consistent at
the quantum level. The ``classical'' approach in dealing with the evidence that extra-dimensions
are not visible up to the sensitivity we have reach so far in experiments, is to assume that they are
compact and very small (of the order of the Plank length), so that any effect due to their presence
could only be seen at energy scales that are completely out of range.

This picture has been modified after the work of Randall and Sundrum \cite{Randall:1999vf}
(a proposal in this direction was made previously by Arkani-Hamed, Dimopoulos and Dvali
in \cite{Arkani-Hamed:1998rs}). In this fundamental paper it was pointed out that current observational
data are not in contrast with the possibility to have an infinite extra-dimension. In fact,
if the universe were made of a 4-dimensional brane embedded in a 5-dimensional $AdS$ bulk,
gravitational interactions should behave as newtonian gravity,
for an observer on the brane, up to small distances (the gauge interactions can be confined
on the brane by a mechanism known in string theory \cite{Horava:1995qa}).

This striking result has inspired a huge number of papers, devoted to develop a cosmology on the
brane that is compatible with current observational tests \cite{Binetruy:1999ut,Shiromizu:1999wj}.
On the other side there were much efforts in finding phenomenological consequences that could in principle
allow us to distinguish between braneworld and standard cosmological evolution.
The most investigated topic for this purpose is the study of the evolution of cosmological
fluctuations that should be produced and amplified during inflation, even in braneworld models
\cite{Langlois:2000ns,Koyama:2000cc,Bridgman:2001mc,Easther:2003re,Kobayashi:2003cn} (for a
detailed review of the main topics covered in these years speculating on the braneworld scenario,
see \cite{Maartens:2003tw}).

In this talk we present a simple model, discussed in \cite{Cavaglia:2005id},
in which a higher-dimensional brane with an anisotropic geometry is embedded in an AdS bulk.
We assume that there are no matter sources on the brane, and the inflationary evolution is
supported by the contraction of internal dimensions. This allow us to decouple the
time-dependent brane evolution equations from the bulk equation, and
this equation itself can be analyzed by using the mode expansion as in ordinary RS case.
We find that in our case there is no mass gap between the massless and the massive modes, and the
amplification of the massless mode is controlled, as in standard cosmology, by the brane curvature.
On the other side, we find that light massive modes are also amplified during inflation,
and under particular circumstances this may lead to a significant enhancement of the final background.

We consider a geometry in which a $(p+1)$-dimensional brane is embedded in a $(p+2)$-dimensional
bulk, with the extra spatial dimension being infinite, and with a negative bulk cosmological constant.
The action describing this system can be put in the form (of course we put $p+2=D$):
\bea
S~=~S_{\rm bulk} + S_{\rm brane}&=& - \frac{1}{2}M^p \int d^{D} x\sqrt{\left|g \right|}
\left(R+2\Lambda \right) \nn \\
& & -\frac{T_p}{2} \int d^{p+1}\xi
\sqrt{\left| \g \right|} \left[\g^{\a \b}
\de_\a X^A  \de_\b X^B g_{AB}(X) - (p-1) \right]
\label{Action}
\eea
where $\L$ is the negative bulk cosmological constant, $T_p$ is the
tension of the brane, and $M$ is a mass parameter that characterizes the
strength of bulk gravitational interactions. The spacetime coordinates are denoted by
$x^A=(t,x^i,y^a,z)$ where $i$ runs from 1 to $d$ and $a$ from 1 to $n$ (the complete brane coordinates
will be denoted with $x^\a$, with $1<\a<p=d+n$), and $z$ is the infinite extra-dimension.
The action is written in terms of the parametrical embedding of the brane coordinates $X^A$ into the
bulk, and of the auxiliary field $\g_{\a \b}$ that will represent the induced metric on the brane.
This formalism can easily be proved to be equivalent to the one proposed in \cite{Shiromizu:1999wj},
at least under the assumption that there is no matter on the brane.
The covariant Einstein equations can be straightforwardly obtained by variating the action
(\ref{Action}) with respect to the bulk metric $g_{AB}$, and variation with respect to the
brane coordinates $X^A(\xi)$ and to the auxiliary metric $\g_{\a \b}$ gives respectively the
brane equation of motion and the constraint that relate the brane metric to the bulk one.

We seek for a solution of the form:
\be
ds^2 = f^2(z)\left[ dt^2 - a^2(t) {\bf dx}^2 - b^2(t) {\bf dy}^2 - dz^2 \right]
\label{metricansatz}
\ee
keeping fixed the brane at $z=0$. (we use directly the conformal frame for the $D$-dimensional metric)
The induced metric takes the expected form, $\g_{\a \b} = \left. g_{\a \b} \right|_{z=0}$,
and the Einstein equations become:
\bea
&& -p F' -\frac{p(p-1)}{2} F^2 + \frac{d(d-1)1}{2}H^2 + \frac{n(n-1)1}{2}G^2 + dn HG = \Lambda_D f^2
\label{Einst00} \\
&& - p F' - \frac{p(p-1)}{2} F^2  + (d-1) \dot H + \frac{d(d-1)}{2} H^2 + n \dot G +
\frac{n(n+1)}{2}G^2 +(d-1)nHG  = \Lambda_D f^2 \nn \\
\label{Einstij} \\
&& -p F' -\frac{p(p-1)}{2} F^2 + d \dot H + \frac{d(d+1)}{2}H^2 + (n-1) \dot G
+ \frac{n(n-1)}{2}G^2 + d(n-1)H G = \Lambda_D f^2 \nn \\
\label{Einstab} \\
&& -\frac{p(p+1)}{2}F^2 + d \left(\dot H + \frac{d+1}{2}H^2 \right) + n\left(\dot G
+ \frac{n+1}{2}G^2 \right)
+dn HG = \Lambda_D f^2
\label{Einst55}
\eea
where dots (primes) denote differentiation w.r.t.\ $t$ ($z$), and $H= \dot a/a$, $G= \dot b /b$, $F=f'/f$.
The time dependent functions that solve the equations have the form:
\bea
a(t) = \left(\frac{t}{t_0}\right)^{\l}\, &~~~~~~& b(t) = \left(\frac{t}{t_0}\right)^{\mu}\ \nn \\
\l = \frac{1 \pm  \sqrt{\frac{n(d+n-1)}{d}}}{d+n}\, &~~~~~~& \mu =
\frac{1 \mp  \sqrt{\frac{d(d+n-1)}{n}}}{d+n}\,
\label{tsol}
\eea
and the $z$-dependent function that gives the $AdS$ warp factor is:
\be
f(z) = \left(1+\frac{|z|}{z_0}\right)^{-1},~~~~~~~~z_0 = \sqrt{-\frac{p(p+1)}{2\Lambda}}
\label{warpfactor}
\ee
provided there is a relation between the brane tension, the mass parameter and the bulk cosmological
constant, which result in tuning to $0$ the brane cosmological constant:
\be
\frac{2T_p}{M^p}= \sqrt{-\frac{32p\Lambda}{p+1}}.
\label{finetuning}
\ee

Now we will discuss tensor perturbations around the background found above. We will consider
a transverse and traceless perturbation $h_{AB}$ of the background metric (\ref{metricansatz})
which takes values  only on the external dimensions ($h_{AB} \equiv h_{ij}$) and depends on $x^i$ and $z$.
The equation of motion for the perturbations is the $D$-dimensional d'alembertian in $AdS$ space:
\be
\ddot h +\left(dH +n G\right) \dot h -\frac{\nabla^2}{a^2}h - h'' - pFh' = 0
\label{perteq}
\ee
This equation is valid for each of the two polarization modes of the gravitational wave, so we
omit the tensor indices. This equation can also be obtained by varying the perturbed (to the
second order) action:
\be
\d^{(2)} S_{(a)} = \frac{M^p}{4} \int d^Dx a^d b^n f^p \left[
\dot h^2 - \sum_{i=1}^d \frac{(\de_i h)^2}{a^2} -h^{\prime 2}\right]
\label{pertaction}
\ee
The variational approach will be necessary when we will try to normalize the perturbations
to an initial state of vacuum fluctuations. Now, following \cite{Randall:1999vf}, we will expand
the perturbation in eigenmodes of the extra-dimension
\be
h(t,x^i,z)= v_0(t,x^i)\psi_0(z) + \int dm~v_m(t, x^i) \psi_m(z) = h_0 (t,x^i,z) + \int dm~h_m(t,x^i,z)
\label{pertexp}
\ee
and, substituting this expansion in the perturbations equation, we get the decoupled equations for the
bulk and the brane modes:
\bea
\psi''_m +pF\psi'_m = - m^2 \psi_m \label{pertsep1} \\
\ddot v_m +\left( dH +nG \right) \dot v_m -\frac{\nabla^2}{a^2} v_m = - m^2 v_m.
\label{pertsep2}
\eea
Defining the auxiliary fields $\psih_m=f^{p/2}\psi_m$ (with a multiplicative factor $\sqrt{M}$ for
the zero mode) allow us to write eq. (\ref{pertsep1}) as a Schr\"{o}dinger-like equation
\be
\psih''_m + \left[m^2 - \left(\frac{p^2}{4} F^2 + \frac{p}{2} F'\right) \right] \psih_m = 0
\label{auxperteq}
\ee
which can be normalized in the canonical way \cite{Randall:1999vf,Maartens:2003tw,Bozza:2001xt,Csaki:2000fc}
\be
\int_{-\infty}^{+\infty} dz~ \psih_m \psih_{m'} = \d(m,m')
\label{norm1}
\ee
where $\d(m,m')$ denotes a Kronecker symbol for the discrete part of the
spectrum and a Dirac distribution for the continuous one.
Substituting the mode expansion (\ref{pertexp}) and the normalization (\ref{norm1})
into the action (\ref{pertaction}) we obtain that the complete action can be written as an infinite sum
over all the mode actions, which are decoupled. The expression of the single mode action, written
in terms of the effective perturbation on the brane $\bar{h}_m = \left. h_m \right|_{z=0}$ is
\be
\d^{(2)} S_m = \frac{M^d}{4\psi^2_m(0)} \int d^{d+1}x~a^d(t) b^n(t) \left[
\dot{\bar h_m^2} - \sum_{k=1}^d \frac{(\de_k \bar h_m)^2}{a^2} -m^2 \bar h_m^2\right]
\label{4Dpertaction}
\ee

Now we have a decoupled action, so we can use the standard technology to obtain the spectrum of
the gravitational fluctuations \cite{Mukhanov:1990me}. We consider an hypothetical phase transition
between the inflationary phase described so far and a standard Minkowsky phase.
We describe cosmological evolution with conformal time $d\eta = dt/a$ and diagonalize the action
introducing an auxiliary field $u_m(\eta)$ via the pump field $\xi_m(\eta)$
\be
\xi_m(\eta) = \sqrt{\frac{M^d}{2}}\frac{1}{\psi_m(0)} a^{\frac{d-1}{2}}b^{\frac{n}{2}}~~~~~~~~~
u_m(\eta,x^i) = \xi_m(\eta)\bar h_m(\eta,x^i)
\label{canvar}
\ee
The resulting canonical action describes a canonical scalar field with a potential which depends on
the background geometry (the zero mode action is not multiplied by the factor $1/M$)
\be
\d^{(2)} S_m = \frac{1}{2M} \int d^dx d\eta~\left[
u_m^{'2} - \sum_{k=1}^d (\de_k u_m)^2+\left(\frac{\xi_m''}{\xi_m} - m^2a^2\right) u_m^2\right]
\label{4Dconfaction}
\ee
From this we get the canonical evolution equation for the Fourier modes $u_{m,k}$ of the
canonical field:
\be
u_{m,k}'' + \left[ k^2+ m^2a^2 - \frac{\xi_m''}{\xi_m}\right] u_{m,k}=0.
\label{modeequation}
\ee
The effective potential of this equation vanishes as $\eta \rarr -\infty$, so
we can normalize the modes to an initial state of vacuum fluctuations
$u_{m,k}(\eta \rarr -\infty)=\exp(-ik\eta)/\sqrt{2k}$. This allows us to deduce the correct
normalization for the solution of (\ref{modeequation}) at all times. Then, by using the solution of
(\ref{auxperteq}) we could come back to the effective fluctuation on the brane and evaluate the
spectral amplitude $\d(k) = k^d \left| h_{m,k}(\eta_1) \right|^2$.

Let us begin with solving equations (\ref{auxperteq}) and (\ref{modeequation}) for $m=0$
The properly normalized solutions are:
\bea
\psi_0(z) &=& \sqrt{\frac{p-1}{2Mz_0}} \nn \\
u_{0,k}(\eta) &=& \sqrt{\frac{\pi}{4} |\eta|} H_0^{(1)}(k|\eta|)\
\label{0modesolutions}
\eea
where $H_0^{(1)}$ is the Hankel function of the first kind, and the spectral amplitude derived from these
solutions is
\be
|\d_0(k)|^2=\left(\frac{H_1}{M_P}\right)^{d-1} \left(\frac{k}{k_1}\right)^d \log^2 \frac{k}{k_1}
\label{0spectraldistr}
\ee
Here $H_1 = k_1 /a$ is the curvature scale of the brane at the transition epoch $\eta_1$, $\Mp$
is the effective Plank mass on the brane, as can be deduced from the action (\ref{4Dpertaction}) and
$k_1$ is the ultraviolet cutoff (frequencies higher than $\eta_1$ are not significantly amplified).
This result is exactly the same we would get in inflationary models
with Kaluza-Klein compactification of extra-dimensions.

Now we turn to solve (\ref{auxperteq}) and (\ref{modeequation}) in the case $m \neq 0$. The properly
normalized solution for the extradimensional equation is:
\be
\psih_m(z) = \sqrt{\frac{m(z+z_0)}{2}} \frac{\left[Y_{\frac{p-1}{2}}(mz_0)
J_{\frac{p+1}{2}}(m(z+z_0))- J_{\frac{p-1}{2}}(mz_0) Y_{\frac{p+1}{2}}(m(z+z_0))\right]}
{{\sqrt{J_{\frac{p-1}{2}}^2(mz_0)+Y_{\frac{p-1}{2}}^2(mz_0)}}}
\label{msol}
\ee
On the contrary, it is not easy to find an exact solution of (\ref{modeequation}). We therefore consider
only the contribution of modes lighter than the curvature scale (the analysis of a general case is
left for a forthcoming paper \cite{futuroderisi}). If $m << H_1$, we can ignore the mass dependent term
in (\ref{modeequation}), and obtain an equation similar to the one studied in the massless case,
with a different normalization of the pump field. The spectral amplitude for the single massive mode can
be written as:
\be
|\d_m(k)|^2=\left(\frac{H_1}{M}\right)^{d-1} \frac{|\psi_m(0)|^2}{M}
\left(\frac{k}{k_1}\right)^d \log^2 \frac{k}{k_1}
\label{massspectraldistr}
\ee
Integration over $m$ leads then to the final spectrum:
be written as:
\be
|\d(k)|^2=\left(\frac{H_1}{M_*}\right)^{d-1} \left(\frac{k}{k_1}\right)^d \log^2 \frac{k}{k_1}
\label{spectraldistr}
\ee
where
\be
M^{d-1}=\frac{M^{d-1}}{\int^{H_1}\frac{dm}{M}|\psi_m(0)|^2}
\label{Mstar}
\ee
We thus obtain again the distribution (\ref{0spectraldistr}), but this time the spectrum is
controlled by an effective mass parameter $M_*$. So, in order to discuss the importance of this contribution
we must estimate the integral in (\ref{Mstar}). If the curvature at the transition epoch is smaller
than the $AdS$ curvature (as we might expect), then $M_* >> M$ and the massive modes contribution
to the spectrum is highly suppressed. On the contrary, if the transition occurs
at very high curvature scales, we obtain that $M_*<M$ and consequently the massive mode contribution
could become relevant, and maybe observable. This amplification could be dangerous for the stability
of the model, because the perturbations could eventually grow until they would destroy  the homogeneity
of the background spacetime. The inflationary model should therefore be completed with an
appropriate mechanism which should stop inflation at a sufficiently low curvature scale.

We are pleased to thank M. Cavagli\`a and M. Gasperini for the collaboration in doing
the paper on which this talk is based, and the organizers of the QG05 conference.

\end{document}